\documentstyle[sprocl]{article}

\def\be{\begin{equation}}
\def\ee{\end{equation}}
\def\bea{\begin{eqnarray}}
\def\eea{\end{eqnarray}}
\begin{document}

%To Prof Nick Karayiannis -- do read this:-
%If needed the word of Chapter~1, you can type in at the 
%\title{}. The words will be in caps and lowercase. 
%For chapter title can be in all caps or in caps and lowercase.
%It is up to the author to type for the case sensitive but 
%all articles must be in the same style. 
%But mostly for Review Volume are without this Chapter~1.
%Thank you
%Jessie   13/4/2000

\title{Gauge Five Brane Moduli in Four Dimensional Heterotic M-theory.}

\author{JAMES GRAY}

\address{University of Durham, Science Laboratories, 
South Road,\\ Durham DH1 3LE, United Kingdom.\\E-mail: j.a.gray2@dur.ac.uk}


\maketitle\abstracts{We present the first example of a K\"{a}hler 
potential for heterotic M-theory which includes gauge bundle moduli. These moduli describe 
the background
 gauge field configurations living on the orbifold fixed planes. We concentrate on the bundle
moduli describing the size and $SU(2)$ orientation of a gauge five 
brane -
 a soliton which is primarily composed of these gauge fields. Our results are valid when 
the width of this object is small compared to the overall size of the 
Calabi-Yau threefold. We find that, in general, it is not consistent to 
truncate away these moduli in a simple manner.}

\section{Introduction}

Heterotic M-theory \cite{Witten:1996mz,Lukas:1998fg} is one of the most promising corners 
of the M-theory 
moduli space studied to date from a phenomenological point of view. The theory
describes eleven dimensional M-theory compactified on an orbifold $\frac{S^1}{Z_2} \times X$ 
where 
$X$ is a manifold of $SU(3)$ holonomy. This compactification
 combines the phenomenological successes of the weakly 
coupled $E8 \times E8$ heterotic string, compactified on a Calabi-Yau threefold, 
with a natural mechanism for obtaining the 
correct strength of both the gravitational and gauge interactions simultaneously.

The vacua of heterotic M-theory are 
non-trivial background solutions which include two key features 
\cite{Witten:1996mz,Lukas:1998fg,Lukas:1999yy}. These are a 
warping of the bulk fields in the orbifold direction and the 
presence of non trivial gauge field expectation values on the two orbifold fixed planes. 

Due to the 
phenomenological success of the theory many authors have naturally been lead to 
study the moduli evolution of its vacua. However no one has been able to include the moduli 
describing the background section of the gauge bundle in these dynamical 
analyses due to the simple fact that the kinetic terms for these fields have 
not been known (although some superpotentials have been computed for such moduli 
\cite{Buchbinder:2002ic,Buchbinder:2002pr}).

The reason for the lack of a four dimensional theory describing these moduli in 
the literature is easy to understand. The most direct way in which to calculate 
the required four dimensional kinetic terms would be to perform a standard dimensional 
reduction about a solution
 describing the vacuum, including the gauge field expectation values. 
However, such a vacuum solution is simply not 
known explicitly.

The aim of this talk is to present the first example of a four dimensional 
theory which includes the kinetic terms for some of these elusive bundle moduli. These results 
were derived in a recent paper \cite{Gray:2003vw}. We 
shall focus on the moduli describing a particular type of gauge field configuration
 known as a gauge five brane. This is a six dimensional solitonic 
object which is essentially a Yang Mills instanton - a localised lump of 
gauge field - dressed up with various other fields.

The plan of this talk is as follows. In section 2 we briefly outline the method which has been 
used to obtain the four dimensional effective action including the gauge five brane moduli. The 
action itself is presented, and its physical consequences discussed, in section 3. Finally some
 promising directions for future research are described in section 4.

\section{Obtaining the effective action}

In deriving the effective theory describing the gauge five brane moduli in 
four dimensions we have to get around the fact that a full background solution describing 
the embedding of the soliton 
within the vacuum is not known. The way we have circumvented this deficiency is 
based upon the fact that the soliton is a localised object in its transverse 
space \cite{Gray:2003vw}.

In any given vacuum configuration the gauge five brane we are interested in must be 
oriented in a specific manner in order to preserve $N=1$ supersymmetry in four dimensions 
\cite{Witten:1996mz}. 
Four of its six worldvolume directions must span the external 
Minkowski space. The remaining two world volume directions wrap a holomorphic 
curve within the Calabi-Yau leaving five directions within the compactification 
manifold, including the orbifold direction, to make up the space transverse to 
the object.
The localised nature of the gauge five brane means that, in some sense, it does 
not probe the compactification manifold in the directions transverse to it's 
world volume at distances which are much greater than the width of its core.
As we shall see this reduces the requirement in our calculation for a solution describing the
entire vacuum to a need to find a solution which is valid
only near to the branes world volume. This is a much simpler task 
as we will now describe.

\subsection{Wrapping up the gauge five brane}

The calculation is most easily preformed within 
the ten dimensional effective description of heterotic M-theory 
in which the orbifold direction has been integrated out \cite{Lukas:1998ew}.
 Using this effective theory as our 
starting point also means that our results are equally valid for the weakly 
coupled heterotic string theories when we make appropriate modifications to the 
numerical values of certain parameters and a suitable modification of the gauge 
group in the $SO(32)$ case. In this talk we shall continue to describe 
everything in the strongly coupled language however for simplicity and unity of 
presentation.

The solution describing a gauge five brane in ten dimensional asymptotically flat 
space is well known \cite{Strominger:et}. 
We need to wrap this uncompactified configuration up in a suitable manner 
on a 2 cycle in a compact manifold. This can be achieved by wrapping up the brane on a Calabi-Yau 
threefold constructed 
as the resolution of an appropriate 
orbifold. In our recent paper 
\cite{Gray:2003vw} a six dimensional orbifold was used with the property that (in 
addition to having $SU(3)$ holonomy) near the Riemann surface, ${\cal C}_2$, on which we are 
going to wrap our gauge 
five brane the compact 
space may be written as $X = {\cal C}_2 \times {\cal C}_4$, where
 ${\cal C}_4$ is a complex four dimensional space.

An example of such an orbifold is a $Z_8 - I$ Coexeter orbifold with a $SO(5) 
\times SO(9)$ lattice.
The 
identifications required to turn flat space into this orbifold which are 
associated with the two cycle ${\cal C}_2$ are consistent with 
the gauge five brane solutions 
symmetries and so may be performed trivially. The identifications made in 
the space transverse 
to the 
brane are not consistent with the symmetries of the asymptotically flat solution. However,
 this merely means that we must modify the configuration at large distances from the core of 
the object and we shall now see that we do not require a solution which is valid in these 
regions.

Given this solution (valid near to the gauge five branes world volume) 
for the background configuration one may then proceed to 
dimensionally reduce in the usual way. We promote integration constants to be four dimensional 
fields and integrate over the compactified extra dimensions in order to obtain a four 
dimensional effective action. If it is indeed the case that 
we only require a background solution for the gauge five brane which is valid near to the 
objects world volume then we should find that the result we get is independent of the gauge 
five branes contributions to the vacuum far out in the transverse space. We find that this is 
indeed the case, thus justifying our approach. The level of approximation involved is determined 
by a well controlled expansion which is described in the next section.

In fact a number of subtleties arise in performing the dimensional reduction which we have not 
mentioned here. These subtleties are described in 
detail in our recent paper \cite{Gray:2003vw}.

\section{The four dimensional action, K\"{a}hler potential and complex structure}

The four dimensional effective action we obtain from the procedure outlined in the previous
section is as follows \cite{Gray:2003vw}.

\begin{eqnarray}
\label{modaction}
S &=& \frac{1}{2 \kappa_4^2} \int d^4x \sqrt{-g} \left( -R + \frac{1}{2} (\partial \varphi)^2 + 
\frac{1}{2}\left( (\partial \beta_1)^2 + (\partial \beta_2)^2 +(\partial \beta_3)^2\right) 
  \right. \\ \nonumber && \;\;\;\;\;\;\; \left. 
+ e^{-2  \varphi} (\partial \sigma )^2 + \frac{2}{9} \left(e^{-2 \beta_1}( \partial \chi_1)^2 +
 e^{-2 \beta_2}( \partial \chi_2)^2 
+ e^{-2 \beta_3}( \partial \chi_3)^2 \right) \right. \\  \nonumber  && \;\;\;\;\;\;\; \left. 
 +  q_{G5}  \left[ \frac{}{} 8 e^{-\beta_1 -\beta_2} \left( ( \partial \rho)^2 + \rho^2 (\partial 
\theta)^2 \right)  + \rho^2 e^{-\beta_1- \beta_2} \left( ( \partial \beta_1 )^2 
+( \partial \beta_2)^2 \right)  \right. \right. \\ \nonumber &&\;\;\;  \left. 
\left.   -4 \rho e^{-\beta_1-\beta_2} \partial \rho \left( \partial \beta_1 + \partial \beta_2 
\right) -\frac{2}{9} \rho^2 e^{-\beta_1 - \beta_2}\left( e^{-2 \beta_1}( \partial \chi_1)^2 
+ e^{-2 \beta_2}( \partial \chi_2)^2 \right)  \right. \right. \\ \nonumber &&\;\;\;\;\;\;\;\;  
\left. \left.  + \frac{4}{9} \rho^2 e^{-2 \beta_1 -2 \beta_2} \partial \chi_1 \partial \chi_2
 \right. \right. \\ \nonumber &&\;\;\;  
\left. \left. +\frac{8}{3} \rho^2 e^{-\beta_1- \beta_2} (\theta^2 \partial \theta^1 -\theta^1 
\partial \theta^2 
+\theta^3 \partial \theta^4 - \theta^4 \partial \theta^3)(e^{- \beta_1} \partial \chi_1 
+ e^{- \beta_2 } \partial \chi_2)   \right] \right)
\end{eqnarray}

This expression employs the following definition.

\begin{eqnarray}
(\partial \theta)^2 = \sum_{\gamma=1}^4 (\partial \theta^{\gamma})^2
\end{eqnarray}

In this result we have the following fields in addition to the four dimensional metric. Firstly 
we have the familiar moduli fields of four dimensional heterotic M-theory \cite{Lukas:1998fg}. 
The 
dilaton field is denoted by $\varphi$ and its associated axion is $\sigma$. The metric moduli describing 
our Calabi-Yau manifold are $\beta_i$ where $i=1,2,3$. The size of the 2 cycle which the gauge 
five brane wraps is determined by $\beta_3$ whereas $\beta_1$ and $\beta_2$ are associated with
 the 
transverse space. The axions associated with these fields 
are $\chi_i$. The moduli describing the blow ups of the orbifold fixed loci have been
truncated. 

In addition to these geometrical moduli we have the bundle moduli 
describing the gauge five brane. These are $\rho$, which describes the solitons width, and 
$\theta^{\gamma}$ where $\gamma=1 ..4$ and $\sum_{\gamma =1}^4 (\theta^{\gamma})^2 = 1$, which 
describe the orientation of the objects gauge field core within $SU(2)$. 
We have also introduced $q_{G5} = \alpha' (2 \pi)^2 / V_{t}$ and 
$V_{t}$ is the coordinate volume of the transverse space to the five brane.

The approximations we have made in obtaining this result are as follows. We have 
made all of the standard approximations employed in obtaining four dimensional actions in this 
context. These are the slowly changing moduli approximation, working to first order in $\alpha'$ 
(or more 
precisely to first order in $\epsilon_w$ in the language used in the literature
 \cite{Lukas:1998hk} ) and ignoring 
towers of 
massive states (which corresponds to 'the other' $\epsilon$ expansion 
\cite{Lukas:1998hk}). The action also does not include contributions from non-perturbative 
corrections.
The new 
approximation that we have made here is that $\rho << V_{t}^{\frac{1}{6}}$ 
(there 
could be corrections to our action which are suppressed by powers of 
$\rho^2 / V_{t}^{\frac{1}{3}}$). In addition, in order for our 
supergravity 
description to
 be valid we require $\rho^2 >> \alpha'$. In other words the gauge five brane has to be wide 
enough to be describable by supergravity but 
narrow enough to be viewed as a localised object. This leaves us with a wide range of five brane 
widths for which our results are valid.

Since the four dimensional action has been constructed to be $N=1$ supersymmetric it can be 
written in terms of a K\"{a}hler potential and an associated complex structure.

\begin{eqnarray}
K &=& -\ln(S + \bar{S}) - \ln(T_1 + \bar{T}_1) - \ln(T_2 + \bar{T}_2) \\ \nonumber && -
\ln(T_3 + \bar{T}_3) 
+ \frac{16 \alpha' \left( |C_1|^2 + |C_2|^2 \right)}{\sqrt{\left( T_1 + \bar{T}_1 \right) 
\left( T_2 + \bar{T}_2 \right)}} \\
C_1 &=& e^{-\frac{\beta_1}{4} - \frac{\beta_2}{4}} \left( Y_1 + i Y_2 \right) \\
C_2 &=& e^{-\frac{\beta_1}{4} - \frac{\beta_2}{4}} \left( Y_3 + i Y_4 \right) \\
T_1 &=& e^{\beta_1} + \frac{2}{3} i \chi_1 + 4 \alpha' e^{\frac{\beta_1- \beta_2}{2}} 
\left( |C_1|^2 + |C_2|^2 \right) \\
T_2 &=& e^{\beta_2} + \frac{2}{3} i \chi_2 + 4 \alpha' e^{\frac{\beta_2- \beta_1}{2}}
 \left( |C_1|^2 + |C_2|^2 \right) \\
T_3 &=& e^{\beta_3} + \frac{2}{3} i \chi_3 \\ 
S &=& e^{\varphi} + \sqrt{2} i \sigma
\end{eqnarray}

Here we have defined the fields $Y_{\gamma}$ as $Y_{\gamma} = \rho \theta^{\gamma}$. 
We see that we have the usual K\"{a}hler potential of heterotic M-theory with an 
additional 
term, the final one, which is due to the presence of the gauge five brane.
 Similarly the definitions of the $T$ 
superfields in terms of  component fields are just the usual ones with a couple of modifications 
at $O(\alpha')$ due to the presence of the gauge five-brane.

We can make a number of comments about the physics that follows from these results
 purely from an examination of the component action and K\"{a}hler structure. 

First of all it is clearly not consistent, in the case of the compactifications
considered here, to take the universal case for the metric moduli in the presence of a
generic changing instanton configuration. In other words, due to the non trivial factors of
$e^{\beta_1}$ and $e^{\beta_2}$ in the gauge five brane moduli kinetic terms, for example, it is
inconsistent to set $\beta_1=\beta_2=\beta_3$ (however as we shall see shortly
 we can set $\beta_1=\beta_2$ if we make some compatible truncations of the other fields).

Secondly it can be seen from the last five terms in equation (\ref{modaction}) that it is not
consistent to
truncate off the gauge five brane moduli by setting them to be non-zero constants. In fact it
is not possible to truncate them away by setting all the $Y$'s to
 zero either, even though they appear bilinearly in the above expressions. This is because setting
 all of the $Y$'s to zero in this manner corresponds to setting $\rho$ to zero and as mentioned
 above our effective description is not valid in this region of moduli space.

More complicated
 forms of truncation are possible in certain special cases. The simplest 
consistent truncation we may obtain of equation (\ref{modaction}) which includes the 
gauge five brane's size modulus, the field we are perhaps most interested in, is given 
below.

\begin{eqnarray}
\label{simplecomponentS}
S = \frac{1}{2 \kappa^2} \int d^4 x \sqrt{- g} \left( -R + (\partial \beta)^2 + 8 q_5 e^{-\beta} 
(\partial
\hat{\rho})^2 \right)
\end{eqnarray}
Here we have taken $\beta_1 = \beta_2 = \beta$,$  \chi_1, \chi_2, \chi_3,
\phi, \sigma, \beta_3$ and $\theta^{\gamma}$ to be zero
 and we have defined $\hat{\rho}  = e^{-\frac{\beta}{2}} \rho$.

This action provides a simple starting point for an investigation of the dynamics associated with 
the bundle modulus $\rho$ and will be at the heart of some future work \cite{glp}.

\section{Future work}

There are many ways in which this work can be used as a basis for future study. Here we will 
describe a
few of the more exciting possibilities.

One could generalise the results presented here to the case where we consider more than one,
possibly overlapping,
gauge five brane. Such complicated
situations are probably tractable due to the fact that we have a very powerful mechanism for
obtaining the self dual gauge field configurations on which such objects are based
 in the form of the ADHM construction
\cite{Dorey:2002ik}. Similar calculations
 to the one presented here could be performed for these more complicated situations.
Indeed by using a Kummer style construction for the Calabi-Yau threefold, such as the one we
 have employed here, and by taking the case where the gauge field background is entirely in
the form of (either overlapping or not overlapping) gauge five branes one could write down a
four dimensional theory which includes {\it all} of the moduli present in the compactification.

There are other possible constituents 
of the gauge bundle which would yield to our approach. For example
there is another localised object which takes a Yang-Mills instanton as its core - the so called 
symmetric
solution \cite{Duff:1994an}. We could equally well apply our method
to this object and obtain the low energy effective theory which includes its moduli.

One could try to obtain a more complete description of the gauge five brane's four dimensional
effective theory by combining the kinetic terms described here with the work which has been done
on obtaining non-perturbative superpotentials for gauge bundle moduli
\cite{Buchbinder:2002ic,Buchbinder:2002pr}. In particular it would be
interesting to identify the moduli in these papers which correspond to those described here, for
example the instanton size.

\vspace{0.2cm}

The action presented in this paper could be used to derive a number of different types of
cosmological solutions. For example one could seek to describe the cosmological effects of a gauge
five brane spreading out with time \cite{glp} or spinning in $SU(2)$ space. Such solutions could
be of critical importance in certain cosmological scenarios 
\cite{Khoury:2001wf,Bastero-Gil:2002hs}.

Our results could be used to improve the description of
cosmological scenarios where the gauge five brane is created as the result of a 
small instanton transition \cite{Khoury:2001wf,Bastero-Gil:2002hs}.
However we would like to stress that these objects can live on the orbifold fixed planes
 irrespective of whether or not the system has undergone such phase transitions. Therefore it is
possible to base scenarios purely on the dynamics of gauge five branes. For example, if we were to
include the position moduli of the five brane in our analysis we could imagine basing some kind
 of brane inflation scenario on gauge five branes and anti gauge five branes. This
soliton-antisoliton inflation could potentially have some quite nice properties. For example
 when inflation ends with the collision of the instanton and anti instanton they would presumably
 annihilate in a manner which is describable within the regime of low energy field theory -
both objects simply being made out of low energy fields 
\footnote{ Although some caution is called for
with this statement given the results presented in the literature
for a situation which one would think would be subject to similar arguments \cite{Lima:1999dn}.
 The two situations
 {\it are} different however. In particular in our case the two colliding objects would have no
net five brane charge.}. The energy from such
 an annihilation could reheat the universe - the fact that the colliding objects are annihilating
 on an orbifold fixed plane presumably means that it would be natural for a sizable proportion of
 the resulting energy to be dumped into matter fields.

In other words gauge five branes can be every bit as useful in developing various scenarios as 
their
'fundamental' counterparts - and in addition these solitonic objects have extra attractive 
features
such as variable widths and the fact that they are entirely built out of low energy fields.

In short it is now possible to start an analysis of the effect of certain types of gauge bundle 
moduli on different cosmological scenarios for the first time.

\section*{Acknowledgements}
JG is supported by a Sir James Knott fellowship and would like to thank the CPT at the 
University of Durham 
for their hospitality while this talk was being written up.

\end{document}